\documentclass[journal=apchd5,manuscript=article]{achemso}
\usepackage{amsmath,amssymb,bm}
\usepackage{graphicx}
\usepackage{booktabs}
\usepackage{xcolor}
\usepackage{microtype}
\usepackage{mdframed}

\author{Ary Portes}
\affiliation{Department of Electronic Engineering, Universidade Federal de Minas Gerais, Belo Horizonte, MG, Brazil}

\author{Narges Dalvand}
\affiliation{Department of Electrical Engineering, École de Technologie Supérieure (ÉTS), Montreal, QC, Canada}

\author{Felipe M. F. Teixeira}
\affiliation{Department of Electronic Engineering, Universidade Federal de Minas Gerais, Belo Horizonte, MG, Brazil}

\author{Omar P Vilela Neto}
\affiliation{Department of Computer Science, Universidade Federal de Minas Gerais, Belo Horizonte, MG, Brazil}

\author{Julian L. Pita Ruiz}
\affiliation{Department of Electrical Engineering, École de Technologie Supérieure (ÉTS), Montreal, QC, Canada}

\author{Michaël Ménard}
\affiliation{Department of Electrical Engineering, École de Technologie Supérieure (ÉTS), Montreal, QC, Canada}

\author{Jhonattan C. Ramirez}
\affiliation{Department of Electronic Engineering, Universidade Federal de Minas Gerais, Belo Horizonte, MG, Brazil}
\email{jcordoba@cpdee.ufmg.br}

\title{Phase-aware Inverse Design for Silicon Photonic Logic Gates}
\abbreviations{SOI, silicon-on-insulator; FoM, figure of merit; FDTD,
finite-difference time-domain; PEC, perfect electric conductor; CR, contrast ratio;
XOR, exclusive-OR; NAND, not-AND; NOR, not-OR; MMA, method of moving asymptotes;
GDS, graphic database system; PC, photonic crystal}
\keywords{photonic logic gates, inverse design, topology optimization, adjoint method,
silicon photonics, XOR gate, interference}

\begin{document}

\section{Abstract}

Inverse design has emerged as a powerful strategy for realizing compact photonic devices; however, its application to logic operations remains constrained by challenges in physical interpretability, architectural generality, and experimental validation. This work introduces an experimentally validated, unified, and physics-driven inverse design framework that implements all fundamental Boolean logic gates within a single silicon photonic architecture. Devices are fabricated within a $2\times2~\mu\mathrm{m}^2$ design region on a silicon-on-insulator platform, representing one of the smallest area reported for photonic logic elements. By integrating amplitude, phase, and energy conservation into a composite figure of merit, the proposed approach enables direct control over constructive and destructive interference. Consequently, all logic functions, including XOR and three-input NAND/NOR operations, are achieved using a standardized configuration with two logical inputs and a bias port. Experimental results exhibit good agreement with numerical simulations across the C-band, confirming both the predictive accuracy and fabrication robustness of the method. Performance benchmarking reveals competitive contrast ratios compared to previous implementations, while providing a unified and scalable design strategy. These findings establish a physically interpretable and experimentally validated paradigm for inverse-designed photonic logic, advancing the development of compact, integrated optical computing systems.

\section{Introduction}
\label{sec:intro}

The rapid expansion of artificial intelligence and the associated increase in data center workloads are driving electronic interconnects toward their fundamental bandwidth and energy limitations~\cite{IEA2024,Miller2017}. Photonics offers a promising alternative by enabling low-loss signal propagation without the resistive heating characteristic of electrical interconnects~\cite{miller2009device}, supporting information transmission across a broad spectrum of wavelengths, and leveraging interference to facilitate direct computation in the optical domain~\cite{Lin2018,Shen2017}. Among photonic technologies, silicon-on-insulator (SOI) platforms are distinguished by their compatibility with CMOS fabrication processes and the high refractive index contrast between silicon ($n \approx 3.48$) and silica ($n \approx 1.44$) at 1550 nm, which enables strong optical confinement in sub-micrometer waveguides. All-optical logic gates implemented on SOI platforms have the potential to execute Boolean operations at the speed of light while significantly reducing energy consumption compared to their electronic counterparts~\cite{Jiao2022, Caballero2022,Pedraza2020}. Considerable research has focused on realizing universal logic operations through various mechanisms, such as plasmonic waveguides with high intensity contrast ratios~\cite{Peng2018}, spatial diffractive neural networks~\cite{Qian2020}, and nonlinear liquid metal platforms~\cite{Xu2024}. However, consolidating a complete, ultra-compact, and experimentally validated library of Boolean logic gates within a single, structurally invariant silicon-photonic architecture remains a fundamental bottleneck.

Recent progress in inverse design, particularly adjoint topology optimization, have transformed the development of compact photonic components~\cite{Molesky2018,Lalau-Keraly2013,Pita2025}. For instance, wavelength demultiplexers~\cite{Piggott2015}, power splitters, and mode converters have been fabricated with minimal footprints, surpassing the limitations of traditional analytical approaches. Several research groups have applied these methods to photonic logic~\cite{Neseli2022,Neseli2025,Wang2024,Lan2024,He2022,Gangaraj2024}. Nevertheless, a critical limitation persists: objective functions in all previous compact adjoint logic designs constrain only amplitude and do not directly address the phase relationships between interfering optical paths~\cite{Neseli2022, Wang2024, Gangaraj2024}.

This work presents an experimentally validated, unified, physics-driven inverse design framework that enables the realization of all fundamental Boolean gates within a single compact photonic architecture. The proposed approach employs a composite objective function that enforces amplitude control, power balance, and phase constraints to regulate optical interference. Using this framework, a comprehensive library of fundamental Boolean gates (OR, AND, XOR, NAND, and NOR) is designed within a common $2 \times 2\,\mu\mathrm{m}^2$ design region patterned in a 220~nm-thick silicon device layer. The resulting devices exhibit spectrally robust responses across the C-band range, logical contrasts aligned with the physical constraints of energy conservation, and operational mechanisms governed by constructive or destructive interference. Experimental validation demonstrates quantitative agreement between simulation and measurement in terms of transmission responses, logical thresholds, and contrast ratios, confirming the optimization framework's robustness to fabrication imperfections. This approach provides a scalable and physically grounded pathway to integrated photonic logic, supporting the progression from individual gates to complex photonic architectures.


\section{Design Framework}

\begin{figure}[H]
\centering
\includegraphics[width=0.9\linewidth]{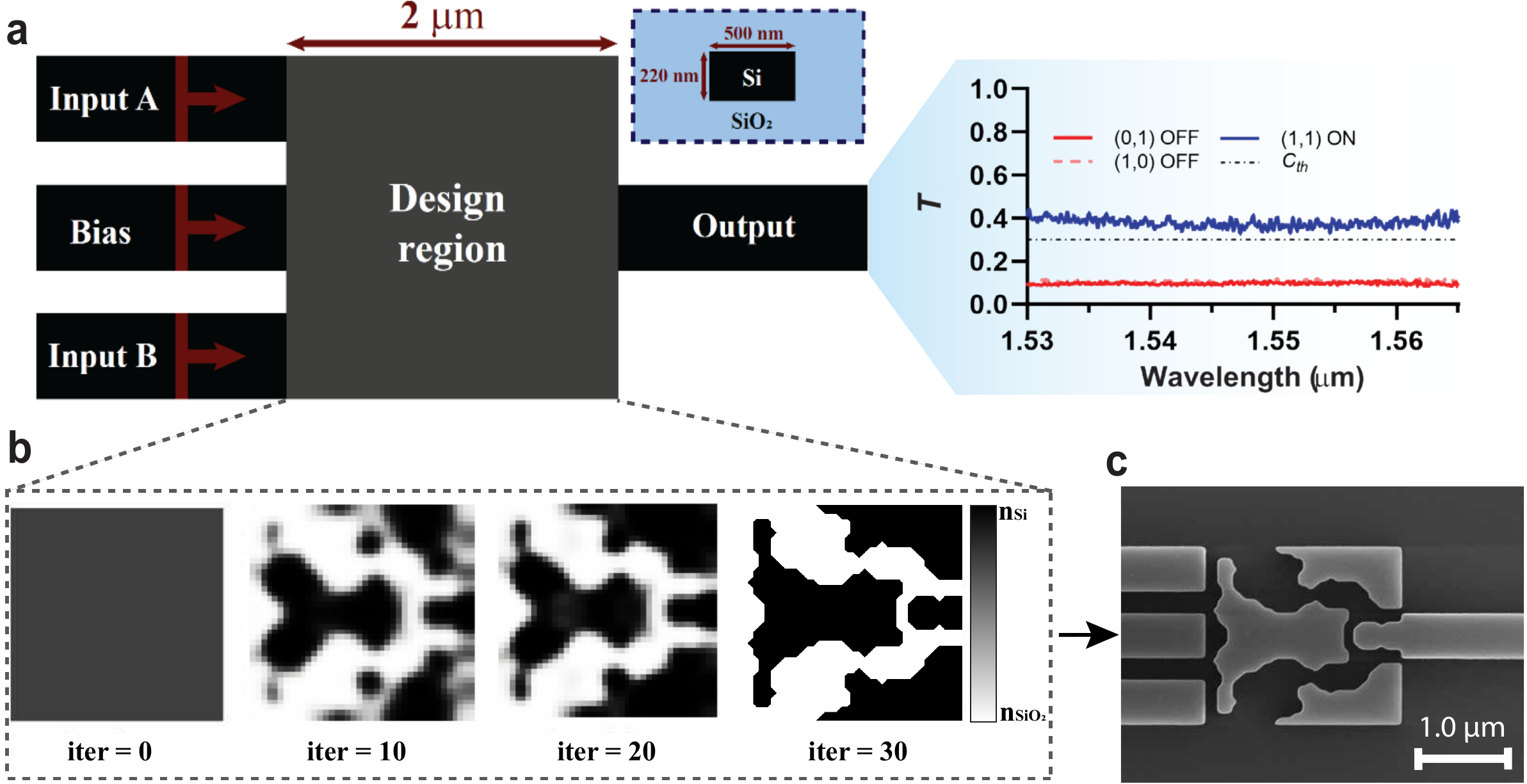}
\caption{
\textbf{a.} Layout of the silicon-on-insulator (SOI) platform highlighting the $2\times2~\mu\mathrm{m}^2$ design region, connecting three 500~nm-wide input waveguides to a single output waveguide. In addition, the broadband transmission spectrum of AND logic gate, showing its response for the input states: (0,1), (1,0), and (1,1), is presented. The inset shows a cross-sectional schematic of the device, consisting of a silicon layer ($n_\mathrm{Si} = 3.48$, $h = 220~\mathrm{nm}$) fully embedded in a silicon dioxide cladding ($n_{\mathrm{SiO}_2} = 1.44$). 
\textbf{b.} Transition from the initial design region to the final device through the inverse design methodology.
\textbf{c.} Scanning electron microscopy (SEM) image of the fabricated adjoint-optimized topology shown in \textbf{b}.
}
\label{fig:overview}
\end{figure}

All five Boolean gates are implemented on a unified silicon photonic platform, as illustrated in Fig.~\ref{fig:overview}a. This SOI wafer layer consists of a $2\times2~\mu$m$^2$ design region on a silicon-on-insulator stack, which is fed by three 500~nm-wide input waveguides and terminated by a single output waveguide. The cross-sectional inset provides details of the silicon layer within its silica cladding. The accompanying broadband spectrum presents the output transmission at each wavelength for every combination of logical inputs, exemplified here for the AND gate. This architectural template is consistent across all gates, maintaining a fixed footprint. Only the topology of the design region varies between logic functions, with its functionality determined by the optimization objective.

The topology that differentiates each gate is determined through topology optimization via the adjoint method. Starting from a uniform design region, the adjoint approach evaluates the gradient of the objective with respect to permittivity at each point using only a forward and an adjoint simulation. This process iteratively guides the continuous material distribution toward a binary distribution that fulfills the target logic (Fig.~\ref{fig:overview}b). The optimized devices were subsequently fabricated on the silicon-on-insulator platform, accurately reproducing the designed topology, as demonstrated by the scanning electron microscopy image in Fig.~\ref{fig:overview}c. Comprehensive details on the optimization procedure and fabrication process are provided in the Methods section.

A composite, physics-informed figure of merit (FoM) forms the foundation of the framework, which is structured around three physically independent degrees of freedom: output transmission accuracy, inter-port power balance, and inter-port phase symmetry. For a device with $N$ input ports, the adjoint optimization minimizes the FoM defined as:
\begin{equation}
J = \sum_{i=1}^{N} \left[ w_1\,|T_i - \hat{T}_i|
+w_2\,|T_i - \bar{T}|
+w_3\,\sigma\,(\cos\Phi_i - 1) \right],
\label{eq:fom}
\end{equation}
where $T_i$ is the transmission of the output mode excited by input port $i$; $\hat{T}_i$ is the truth-table target for port $i$; $\bar{T} = \frac{1}{N}
\sum_{i=1}^{N} T_i$ is the mean individual transmission; $\Phi_i$ is the phase accumulated from input $i$ to the output monitor; and $w_1, w_2, w_3 \geq 0$ are scalar weights. The binary selector $\sigma \in \{+1,-1\}$ sets which interference condition the optimizer pursues: for $\sigma = -1$ the objective drives the accumulated phase toward $\Phi_i = 0$ (constructive interference), while for $\sigma = +1$ it drives $\Phi_i$ toward $\pi$ (destructive
interference). The generality of this form is the key point: gate design reduces to parameter selection within one objective, leaving the optimizer and
adjoint machinery unchanged across all five gates.

The FoM is anchored to the analytical limits of coherent superposition, which set the achievable ON and OFF levels for every gate. The output power of $N$ sources with amplitudes $E_i$ and phases $\phi_i$ is
$P_\mathrm{out} = |\sum_{i=1}^{N} E_i e^{i\phi_i}|^2$. For $N$ co-phased inputs of equal amplitude, each normalized to unit input power, this reaches its maximum
\begin{equation}
P_\mathrm{out}^{\max} = N^2 T_1 ,
\label{eq:constructive}
\end{equation}
where $T_1$ is the transmission of a single source acting alone, while energy
conservation caps the per-port efficiency at
\begin{equation}
T_i \leq \frac{1}{N} .
\label{eq:ceiling}
\end{equation}
Conversely, two equal-amplitude inputs in antiphase ($|\phi_2 - \phi_1| =
\pi$) annihilate the output ($P_\mathrm{out} = 0$). These bounds are
independent of device geometry or optimization strategy, and they define the
physical ceiling against which the measured contrast ratios in Sec.~4 are
interpreted. The full derivation is given in the Supplementary Information
(Sec.~S1).

Within this unified objective, each gate is associated with a distinct interference regime, determined by its targets and the sign of $\sigma$. The constructive gates (OR and AND) align their inputs in phase ($\sigma = -1$) and target the interference maximum described in Eq.~\eqref{eq:constructive}. For the AND gate, a deliberately reduced target ensures that single-input states remain below threshold without explicit penalization. The XOR gate operates in a destructive regime ($\sigma = +1$), enforcing cancellation of the dual-input state. The three-input gates (NAND and NOR) incorporate a permanently switched-on bias port, with its amplitude relative to the logic inputs, expressed as the ratio $T_{g,00}/T_{g,01}$, determining
the OFF condition. The two gates differ only in this ratio. The complete set of per-gate targets, weights, and phase conditions is provided in Table~S1.

\section{Results and Discussion} 
\label{sec:results} 

\begin{figure}[H]
\centering
\includegraphics[width=\linewidth]{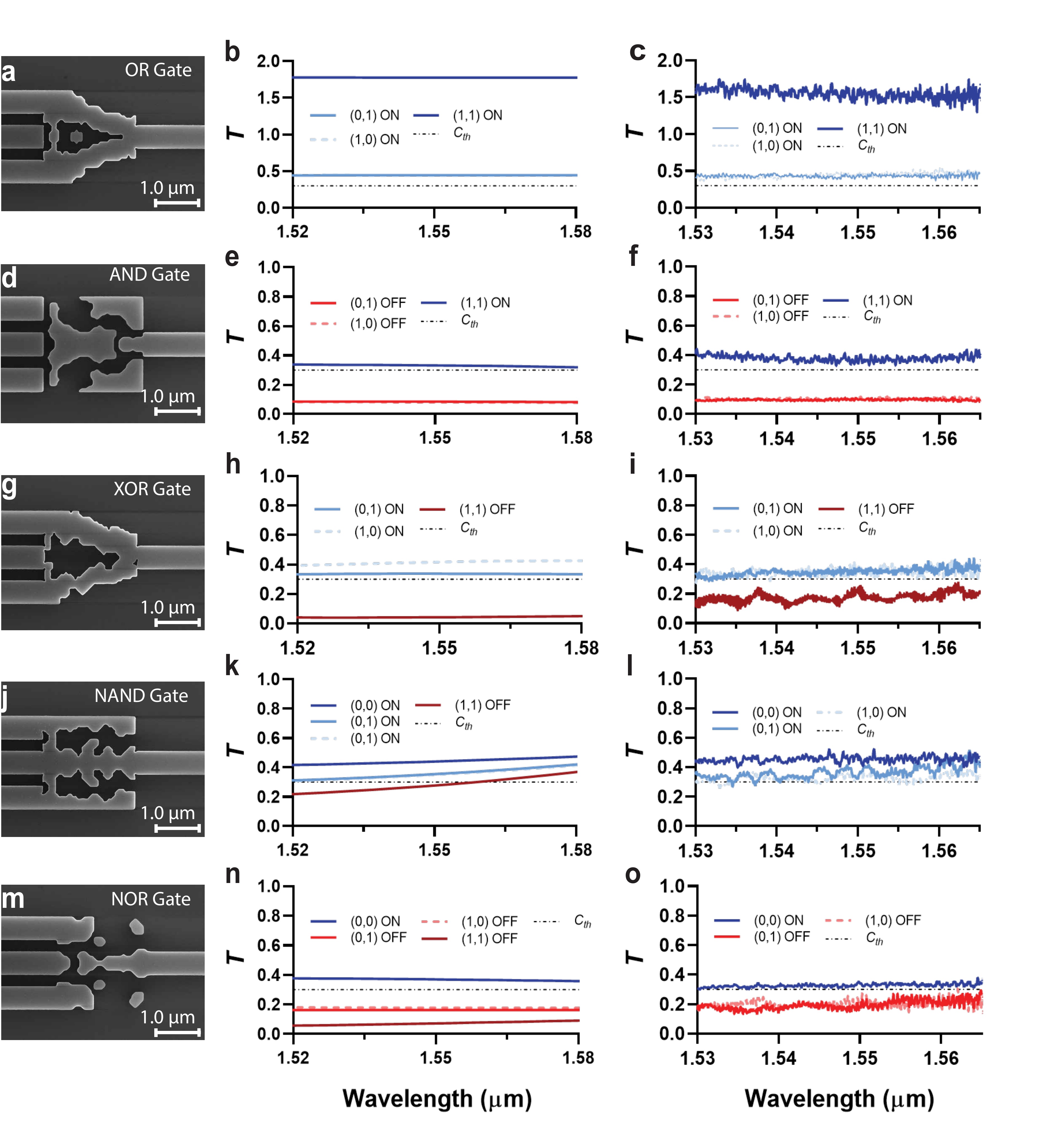}
\caption{
Comprehensive characterization of all-optical logic gates.(\textbf{a, d, g, j, and m}) SEM images of the fabricated OR, AND, XOR, NAND, and NOR devices, respectively, optimized layouts within the $2\times2~\mu\mathrm{m}^2$ design region. Simulated (\textbf{b, e, h, k,} and \textbf{n}) and measured (\textbf{c, f, i, l,} and \textbf{o}) broadband transmission spectra for each logic gate centered at $\lambda = 1550~\mathrm{nm}$).
}
\label{fig:logic_performance}
\end{figure}

Figure~\ref{fig:logic_performance} presents the fabricated devices alongside their simulated and measured transmission spectra. Across the C-band, the measured responses closely align with simulations for each gate, and scanning electron microscopy (SEM) images confirm high-fidelity pattern transfer. The two-input gates illustrate the constructive and destructive interference strategies. In the (0,0) state, the absence of injected light yields no output. For the OR gate, all other input combinations correspond to logical ON states, resulting in the absence of a powered OFF level. Performance is therefore quantified by the margin of the weakest ON state, which remains 1.1~dB above the threshold $C_\mathrm{th}=0.30$ throughout the band. The AND gate is engineered for the same constructive maximum but with a deliberately reduced value, lowering single-input states to $T\approx0.08$ while the dual-input state reaches $0.32$, yielding a 6.0~dB contrast. The XOR gate employs destructive interference: single-input states transmit ($T\approx0.32$), whereas the dual-input state is suppressed to $T\approx0.074$, resulting in a 6.4~dB contrast.

The three-input NAND and NOR gates incorporate a permanent bias that maintains the (0,0) state in the ON condition, while logic inputs interfere with this bias to produce OFF states. The measured ON states are consistent with the design: for NAND, the (0,0), (0,1), and (1,0) states remain above threshold ($T_{00}=0.44$, $T_{01}\approx T_{10}\approx0.35$); for NOR, the (0,0) state remains ON ($T_{00}=0.33$), and a single logic input reduces the output ($T_{01}\approx T_{10}\approx0.15$). 

For NAND and NOR gates, the (1,1) state is determined by the coherent superposition of both logic inputs and the applied bias. Since these devices operate through strictly linear interference, and strong experimental agreement has been established for the orthogonal (0,1) and (1,0) excitations, the (1,1) response is governed by superposition. As a result, the (1,1) OFF levels are accurately reproduced by numerical models, yielding $T_{11}=0.27$ for NAND and $T_{11}\approx0.047$ for NOR, corresponding to contrast ratios of 1.1 and 3.4~dB, respectively.

\begin{figure}[H]
\centering
\includegraphics[width=\linewidth]{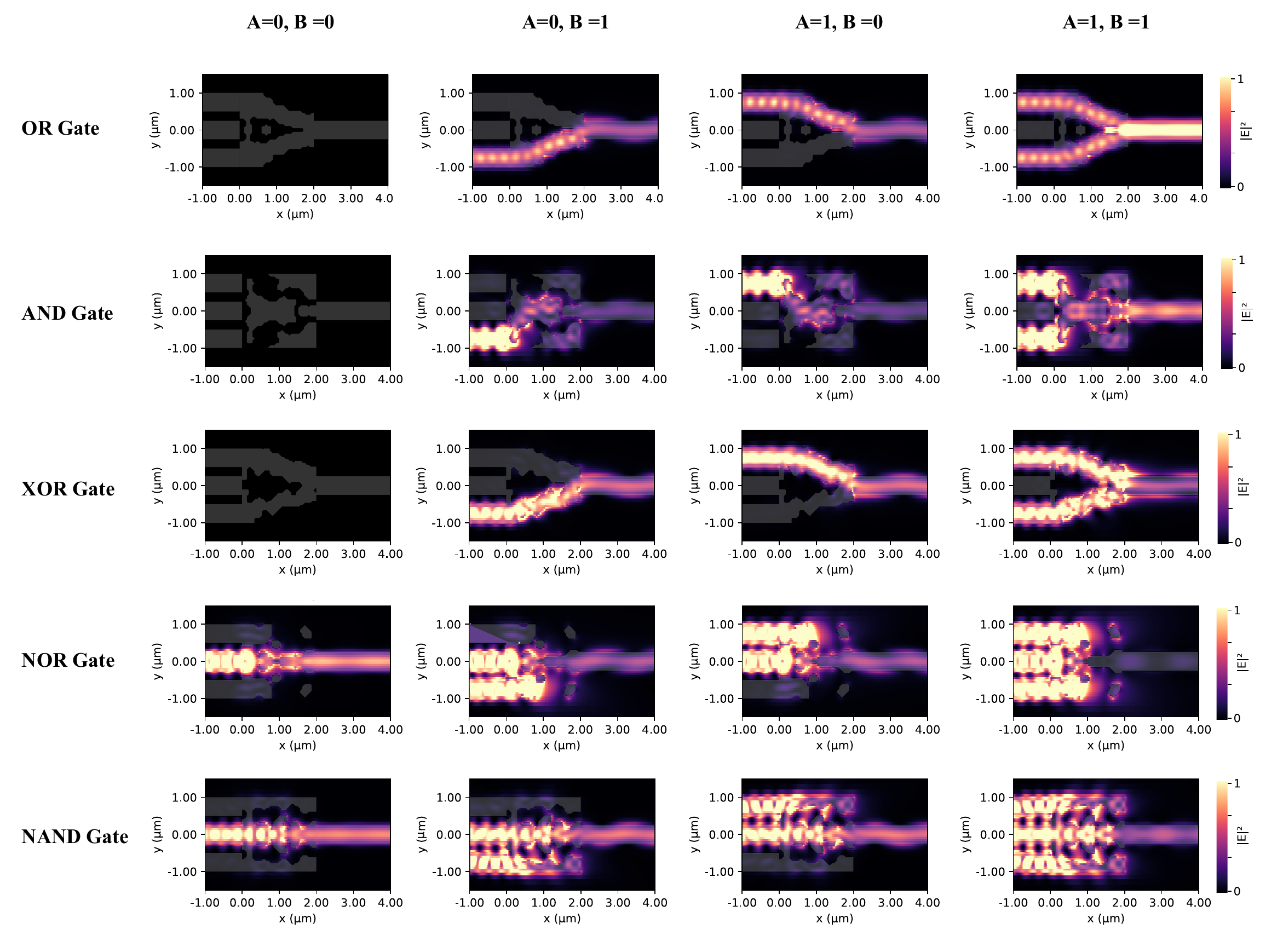}
\caption{
Electric field intensity distributions ($|E|^2$) at $\lambda = 1550~\mathrm{nm}$ for all five logic gates and input combinations. Rows (top to bottom) correspond to OR, AND, XOR, NOR, and NAND gates. Columns (left to right) represent the input states (0,0), (0,1), (1,0), and (1,1). Each row uses a shared colorbar normalized to the 98th percentile of the field-intensity distribution associated with the corresponding gate.
}
\label{fig:fields}
\end{figure}

The electric-field intensity distributions in Fig.~\ref{fig:fields} clarify the operating mechanisms common to all gates. In constructive gates, excited inputs reinforce at the output waveguide, producing a bright output lobe during ON states. For the XOR gate, single-input states direct power to the output, while the dual-input state forms a node at the output monitor, indicating destructive cancellation. In the three-input NAND and NOR gates, the bias field appears as a continuous background.
This background is either reinforced or suppressed by the logic inputs. In all cases, the field distribution adheres to the intended interference pattern rather than exhibiting diffuse scattering, confirming that logical function is encoded through engineered phase relationships among the interfering paths.

Compact photonic logic gates can be classified into two primary design paradigms, distinguished by the location of logic encoding (see Table~\ref{tab:comparison}). In the analytical-interferometry approach~\cite{He2022,Peng2018}, the interference condition is derived analytically and enforced at the sources: each gate requires specific inter-input amplitude and phase offsets maintained by optical phase-locked loops, so part of the logic resides in the inputs. In the amplitude-based approach, which includes inverse-designed~\cite{Neseli2022,Lan2024,Wang2024,wang2025inverse} and photonic-crystal~\cite{Hussein2018,Caballero2022} devices, the logic is embedded in the structure and the inputs are trivial, but the objective constrains only amplitude and allows the interference phase to emerge implicitly.

The present framework occupies an intermediate position: it is fully inverse-designed, with logic encoded entirely in the device topology and all inputs identical in amplitude and phase, achievable from a single source via a splitter. Notably, the interference phase is explicitly incorporated into the objective function. This approach is particularly effective for destructive gates. For example, compared to the closest amplitude-based inverse design~\cite{Neseli2022}, which reports XOR and NAND contrasts of $1.2$ and $0.5$~dB, the phase-aware objective achieves $6.4$ and $1.1$~dB. Achieving higher raw contrast in other approaches often requires trade-offs, such as engineered phase-locked inputs~\cite{He2022,Peng2018}, graded-index designs that are not directly manufacturable and shift wavelength per gate~\cite{Wang2024}, or significantly larger device areas~\cite{Hussein2018,Caballero2022}. The primary contribution is not a record in contrast or footprint, but rather the establishment of a design principle: a single, physically interpretable objective that directly embeds each gate's interference condition into the loss function.

\begin{table*}[h]
\centering
\caption{Comparison of compact photonic logic gate implementations, grouped by design paradigm.}
\label{tab:comparison}
\setlength{\tabcolsep}{5pt}
\renewcommand{\arraystretch}{1.15}
\resizebox{\textwidth}{!}{%
\begin{tabular}{@{}cccccccccc@{}}
  \toprule
  & & & & & \multicolumn{5}{c}{Contrast ratio (dB)$^{a}$} \\
  \cmidrule(lr){6-10}
  Ref. & Approach & Demo. & Size ($\mu$m$^2$) & Band (nm) & OR & AND & XOR & NAND & NOR \\
  \midrule
  \multicolumn{10}{@{}l}{\textit{Analytical interferometry (logic in the inputs)}}\\
  \citenum{He2022} & Analytic + TO & Exp. & 1.7--5.85 & 1520--1600 & 31.8 & 9.5 & 31.8 & 25.8 & 9.5 \\
  \midrule
  \multicolumn{10}{@{}l}{\textit{Amplitude-based design (logic in the structure)}}\\
  \citenum{Neseli2022}      & TO  & Sim. & 5.0          & $\sim$1300 & 5.8          & 4.8  & 1.2  & 0.5  & ---$^{c}$ \\
  \citenum{Neseli2025}      & TO  & Exp. & 30           & 1520--1600 & $\infty^{b}$ & 5.2  & ---  & ---  & --- \\
  \citenum{Lan2024}         & GA  & Sim. & 4.8          & 1550--1600 & 8.6          & 5.3  & 4.14  & ---  & --- \\
  \citenum{Wang2024}        & TO  & Sim. & 4.0          & 800--1600  & 23.5         & 6.1  & 23.5 & 10.1 & 7.9 \\
  \citenum{wang2025inverse} & TO  & Sim. & $\lesssim$30 & 1550       & 4.2          & 4.5 & ---  & 1.1  & 5.5 \\
  \citenum{Hussein2018}     & PhC & Sim. & 123--562     & 1267--1996 & $\infty$     & 6.0  & 11.6 & 5.0  & 8.1 \\
  \citenum{Caballero2022}   & PhC & Sim. & 598--710     & 1550       & 8.7          & 6.2  & 11.6 & 5.7  & 5.9 \\
  \midrule
  \multicolumn{10}{@{}l}{\textit{Phase-aware inverse design (this work)}}\\
  This work & TO & Exp. & 4.0 & 1520--1580 & $\infty$ & 6.0 & 6.4 & 1.1 & 3.4 \\
  \bottomrule
\end{tabular}
}
\vspace{4pt}
{\footnotesize
\noindent
TO: topology optimization; GA: genetic algorithm; PhC: photonic crystal.
$^{a}$~Contrast ratio (CR), defined as $10\log_{10}(P_\mathrm{ON}/P_\mathrm{OFF})$.
$^{b}$~``$\infty$'': gate with no powered OFF state, giving a formally infinite ratio.
$^{c}$~``---'': gate not demonstrated.
}
\end{table*}

Compared to other analytical-interferometry implementations on silicon, such as the topology-optimized gates demonstrated by He et al.~\cite{He2022}, this platform offers distinct advantages in integration simplicity and power efficiency. Whereas the referenced design requires a multi-stage cascade of three separate components, including input Y-junction splitters, waveguide phase-shifting regions, and the optimized logic junction, to form a single functional gate, the present framework monolithically integrates the entire logical operation within a single, invariant $2\times2\,\mu\text{m}^2$ design region. This single-stage architecture eliminates the need for auxiliary routing and splitting components, resulting in devices that are inherently more compact. Additionally, by avoiding the cumulative insertion losses associated with cascading multiple photonic elements, these logic gates are substantially less lossy while maintaining a standardized physical footprint for the entire Boolean library.

The residual discrepancies between simulation and measurement are minor and can be attributed to identifiable physical causes. Fabrication introduces nanoscale variations in feature size, sidewall angle, and material index, which slightly alter optical path lengths and, consequently, the relative phase between interfering fields within the design region. Destructive gates (XOR, NAND, and NOR), which define their OFF states through near-complete $\pi$ cancellation, are inherently more sensitive to these phase perturbations than constructive gates (OR and AND). Therefore, residual deviations are expected to be largest where operation is most phase-critical. This sensitivity is quantified in the Supplementary Information (Fig.~S1), which sweeps the relative phase offset $\Delta\varphi$ of one input across the full $\pm 180^\circ$ range while tracking the output transmission. The phase-sensitive gates exhibit steep extinction-ratio variation with $\Delta\varphi$ and well-defined phase windows within which logical operation is preserved, whereas the constructive gates remain essentially flat. These results indicate that the measured deviations correspond to small, fabrication-scale phase errors rather than uncontrolled scattering.

Robustness to dimensional errors is evaluated in the Supplementary Information (Figs.~S2 and S3). Uniform morphological dilation and erosion of the optimized geometries, simulating over- and under-etching, were applied across a $\pm 20$~nm range, and the transmission of each logic state was re-assessed. Figure~S2 shows that overall topology and connectivity are maintained even at the $\pm 20$~nm extremes, while Fig.~S3 confirms that most gates preserve their logical function throughout the $\pm 10$~nm tolerance typical of advanced silicon photonics. Together, the phase and geometric analyses indicate that the observed simulation-to-measurement deviations are a predictable result of coherent, phase-dependent operation under realistic fabrication tolerances, rather than evidence of uncontrolled scattering.

In summary, the close agreement between simulation and experiment for all five gates validates the predictive capability of the physics-informed framework. Integrating amplitude, power balance, and phase into a unified objective yields compact, broadband, and physically interpretable logic devices whose performance is governed by fundamental interference limits. This approach provides a general and scalable foundation for integrated optical information processing.
\section{Conclusions}
\label{sec:conclusions}

A unified, physics-driven inverse design framework has been developed to synthesize photonic logic within a standardized, compact architectural template. By incorporating interference physics directly into the optimization objective, this method enables precise control over both amplitude and phase at the output. The framework’s interpretability derives from the explicit design objective rather than the resulting topology. In particular, the interference condition required for each gate is predetermined based on energy conservation and coherent superposition, and is subsequently validated against the measured device response. This methodology grounds the design objective in electromagnetic principles, thereby addressing a key limitation of conventional inverse design methods, which often specify objectives without explicit reference to the underlying physics.

The validity and scalability of this framework are demonstrated through the experimental realization of the complete Boolean logic library on a silicon-on-insulator (SOI) platform. All five fundamental logic gates (OR, AND, XOR, NAND, and NOR) are synthesized within an identical, ultra-compact $2\times2\,\mu\text{m}^2$ design region ($4\,\mu\text{m}^2$ total footprint), representing one of the smallest reported footprints for photonic logic elements. Characterization across the C-band ($1520\text{--}1580\,\text{nm}$) reveals excellent quantitative agreement with simulations, yielding competitive measured contrast ratios at $1550\,\text{nm}$ of $6.4\,\text{dB}$ for XOR, $6.0\,\text{dB}$ for AND, $3.4\,\text{dB}$ for NOR, and $1.1\,\text{dB}$ for NAND. The narrower contrast observed in the NAND gate results directly from energy conservation in multi-port coherent superposition, rather than from an optimization limitation. The OR gate demonstrates stable operation with an ON-state transmission margin of $1.1\,\text{dB}$ above the logical threshold ($C_{\text{th}} = 0.30$).

The design pipeline exhibits outstanding resilience to fabrication-induced variations. Even with a strict $100\,\text{nm}$ minimum feature size and spacing constraint to ensure compatibility with standard deep-ultraviolet (DUV) lithography, sensitivity analyses confirm that overall topological connectivity is preserved under uniform morphological dilation and erosion of up to $\pm20\,\text{nm}$. All five gates maintain their logical functionality across this range, exceeding the typical $\pm10\,\text{nm}$ tolerances of state-of-the-art silicon photonics foundries. The minor discrepancies between simulation and measurement are attributed to predictable, nanoscale phase perturbations ($\Delta\phi$) in the phase-critical destructive gates, rather than to uncontrolled diffuse scattering.

In summary, this work establishes a physically grounded and experimentally validated paradigm for inverse-designed photonic logic. By demonstrating that universal logical operations can be encoded within a single invariant footprint with robust fabrication tolerances, these results provide a scalable and highly compact pathway toward next-generation energy-efficient optical computing and information co-processors.

\section{Methods}
\label{sec:methods}

\textbf{Design Optimization: } The inverse design process was implemented using the continuous adjoint method. The design region was initialized with a uniform relative permittivity corresponding to the arithmetic mean of silicon and silica ($\rho_0 = 0.5$), facilitating an efficient symmetry breaking~\cite{Piggott2015}. A dual-step regularization procedure was applied at each iteration: a cone-shaped density filter with a $120$~nm radius to enforce minimum feature sizes compatible with fabrication constraints, followed by a progressive Heaviside projection to ensure a binary, manufacturable topology. The objective function was maximized using the Adam optimizer~\cite{Kingma2014} with a learning rate of $\alpha = 0.01$. Electromagnetic simulations for the forward and adjoint fields were performed using the cloud-based finite-difference time-domain (FDTD) solver Tidy3D~\cite{Tidy3D2024}. Convergence to binary topologies was achieved within 30 iterations for all logic gates. Full mathematical details of the optimization workflow and parameter constraints are provided in the \textit{Supplementary Information}.

\textbf{Fabrication: }The devices were fabricated on an SOI photonic integration platform comprising a 220 nm silicon device layer on a $2~{\mu m}$ buried oxide layer and a $725~{\mu m}$ silicon handle substrate. Fabrication was carried out through a multi-project wafer run at ANT. Device patterns were defined by 100 keV electron-beam lithography (EBL) and transferred into the silicon device layer using a single-step anisotropic inductively coupled plasma reactive-ion etching process. A $2.2~{\mu m}$ silicon dioxide top cladding was subsequently deposited by plasma-enhanced chemical vapor deposition. Although the EBL process supports feature sizes down to approximately 60 nm, a minimum feature size and spacing of 100 nm were adopted for all logic gates to enhance fabrication yield and ensure compatibility with standard 193 nm deep-ultraviolet CMOS manufacturing.

The fabricated waveguides exhibited an average sidewall angle of approximately 88°, and the propagation loss of the fundamental TE mode at 1550 nm was approximately 1.2 dB/cm. Thermo-optic phase shifters for controlling the input optical phase were realized using a two-metal-layer process comprising a high-resistivity titanium–tungsten (TiW) resistive heater and a low-resistance TiW/Au routing layer. The metal layers and a silicon dioxide passivation layer were patterned by photolithography, followed by selective opening of the passivation layer above the contact pads for electrical wire bonding.

\textbf{Characterization: }Optical characterization was performed using an automated EXFO wafer testing station, providing high-precision and repeatable fiber-to-chip alignment. An EXFO T100S-HP tunable laser was used as the optical source, while the transmitted optical power was measured using an EXFO CTP10 optical power meter. For measurements requiring two simultaneous optical inputs, the laser output was split into two paths, each equipped with an independent polarization controller. Light was coupled into and out of the devices through a polarization-maintaining fiber array with an 8° polished angle and surface grating couplers optimized for the fundamental TE polarization.

The insertion losses associated with the grating couplers and routing waveguides were de-embedded using reference loopback structures with optical path lengths matched to those of the corresponding logic gates, thereby isolating the response of the devices under test. The integrated thermo-optic phase shifters were driven using a Keithley 2400 source-measure unit, with applied voltages ranging from 0 to 14 V to control the optical phase during measurements involving two optical inputs. All measurements were performed at room temperature.

\section{Acknowledgements}
\label{sec:acknowledgements}

The authors acknowledge financial support from FAPEMIG (APQ-02286-23), CNPq and CAPES. In addition, they extend their sincere thanks to \textit{Flexcompute} for making the \textit{Tidy3D} software available to them for the numerical/computational analysis performed. The authors also acknowledge the support of CMC Microsystems for the fabrication of the prototypes, of the LaCIME for access to the testing facility, and the financial support of the Natural Sciences and Engineering Research Council of Canada.

\section{Author contributions}
\label{sec:contributions}

\textbf{Idealization and conceptualization:} JCR, OPV, AP. 
\textbf{Simulations and framework development:} AP, FMT, JCR. 
\textbf{Fabrication layout:} AP, FMT, JCR, JPR, MM. 
\textbf{Optical characterization:} ND, JPR, MM. 
\textbf{Data analysis:} AP, JCR, ND, JPR, MM. 
\textbf{Supervision:} JCR, MM, OPV, JPR. 
\textbf{Writing original draft and editing:} All authors.

\bibliography{refs}


\end{document}